\documentclass[twocolumn,showpacs,preprintnumbers,amsmath,amssymb,
superscriptaddress,groupedaddress,floatfix]{revtex4}

\usepackage{graphicx}
\usepackage{dcolumn}
\usepackage{bm}
\usepackage{amssymb}

\begin{document}

\title{Spreading of wave packets in disordered systems with
tunable nonlinearity}

\author{Ch.~Skokos}
\affiliation{Max Planck Institute for the Physics of Complex Systems, 
N\"othnitzer Str. 38, D-01187 Dresden, Germany}
\author{S.~Flach}
\affiliation{Max Planck Institute for the Physics of Complex Systems, 
N\"othnitzer Str. 38, D-01187 Dresden, Germany}

\date{\today}

\begin{abstract}
  We study the spreading of single-site excitations in one-dimensional
  disordered Klein-Gordon chains with tunable nonlinearity
  $|u_{l}|^{\sigma} u_{l}$ for different values of $\sigma$.  We
  perform extensive numerical simulations where wave packets are
  evolved a) without and, b) with dephasing in normal mode
  space. Subdiffusive spreading is observed with the second moment of
  wave packets growing as $t^{\alpha}$. The dependence of the
  numerically computed exponent $\alpha$ on $\sigma$ is in very good
  agreement with our theoretical predictions both for the evolution of
  the wave packet with and without dephasing (for $\sigma \geq 2$ in
  the latter case). We discuss evidence of the existence of a regime
  of strong chaos, and observe destruction of Anderson localization in
  the packet tails for small values of $\sigma$.
\end{abstract}

\maketitle

\section {Introduction}
\label{sec:intro}

The presence of uncorrelated spatial disorder in one-dimensional
linear wave equations results in the localization of their normal
modes (NMs). This is the well-known phenomenon of Anderson
localization \cite{A58} and has been experimentally observed in a
variety of systems, including as examples light
\cite{SBFS07,LAPSMCS08} and matter \cite{BJZBHLCSBA08,REFFFZMMI08}
waves.

Understanding the effect of nonlinearity on the localization
properties of wave packets in disordered systems is a challenging
task, which, has attracted the attention of many researchers in recent
years. Several numerical studies of wave packet propagation in
different models showed that the second moment $m_2$ of norm/energy
distributions grows subdiffusively in time following a power law of
the form $m_2\sim t^{\alpha}$
\cite{M98,KKFA08,PS08,FKS09,GMS09,SKKF09}.

In \cite{FKS09,SKKF09} the mechanisms of spreading and localization
were studied for the disordered discrete nonlinear Schr\"odinger
equation (DNLS)
\begin{equation}
i\dot{\psi_{l}}= \epsilon_{l} \psi_{l}
+\beta |\psi_{l}|^{2}\psi_{l}
-\psi_{l+1} - \psi_{l-1},
\label{DNLS}
\end{equation}
and the quartic Klein-Gordon lattice (KG) of anharmonic oscillators
with nearest neighbor coupling.  The exponent $\alpha$ was numerically
found to be close to $\alpha\approx1/3$ and a theoretical explanation
of this value was provided.

Applying the same theoretical argumentation to a generalized DNLS
(gDNLS) model
\begin{equation}
i\dot{\psi_{l}}= \epsilon_{l} \psi_{l}
+\beta |\psi_{l}|^{\sigma}\psi_{l}
-\psi_{l+1} - \psi_{l-1},
\label{gDNLS}
\end{equation}
with $\sigma$ being a positive integer, the dependence of $\alpha$ on
$\sigma$ was predicted in \cite{FKS09}. The validity of this
estimation has not been analyzed in detail.  Mulansky \cite{M09}
presented numerical simulations of the gDNLS model for a few integer
values of $\sigma$.  In \cite{VKF09} numerical simulations of the
gDNLS model were performed for non integer values of $\sigma$ on
rather short time scales, leaving the characteristics of the
asymptotic ($t \rightarrow \infty$) evolution of wave packets aside
and open.

The main scope of the present work is to verify the validity and the
generality of the theoretical predictions presented in \cite{FKS09}
for one-dimensional disordered nonlinear chains as a function of
$\sigma$. In particular, we choose to perform numerical simulations
for the generalized KG (gKG) model instead of the gDNLS system. This
choice was done for two reasons. Firstly, it allows us to test whether
the estimations obtained in \cite{FKS09} hold irrespectively of the
presence of a second integral of motion (the norm $\sum_l
|\psi_{l}|^2$ for the gDNLS model).  The second reason is a practical
one. From the comparative study of DNLS and KG models (which
correspond to $\sigma=2$) performed in \cite{FKS09,SKKF09} it was
observed that the KG model requires less CPU time than the DNLS system
in order to be integrated up to the same time with the same
precision. Since in our study we are mainly interested in the
characteristics of the asymptotic dynamical behavior of wave packets,
the gKG model was preferred, as it permits long integrations of large
lattice sizes within feasible CPU times.

\section {The generalized Klein-Gordon model}
\label{sec:GKG}

The Hamiltonian of the gKG model is
\begin{equation}
\mathcal{H}= \sum_{l}  \frac{p_{l}^2}{2} +
\frac{\tilde{\epsilon}_{l}}{2} u_{l}^2 + 
\frac{| u_{l} |^{\sigma+2}}{\sigma +2} +\frac{1}{2W}(u_{l+1}-u_l)^2,
\label{H_KG}
\end{equation}
where $l$ is the lattice site index, $u_l$ and $p_l$ are respectively
the generalized coordinates and momenta, $\sigma$ defines the order of
the nonlinearity, $W$ denotes the disorder strength and
$\tilde{\epsilon}_{l}$ are chosen uniformly from the interval
$\left[\frac{1}{2},\frac{3}{2}\right]$. The case $\sigma=2$
corresponds to the standard KG model, which exhibits a similar
dynamical behavior with the DNLS model \cite{FKS09,SKKF09}. The
equations of motion are $\ddot{u}_{l} = - \partial \mathcal{H}
/\partial u_{l}$ and yield
\begin{equation}
\ddot{u}_{l} = - \tilde{\epsilon}_{l}u_{l}
-|u_{l}|^{\sigma} u_{l} + \frac{1}{W} (u_{l+1}+u_{l-1}-2u_l)\;.
\label{H_EM}
\end{equation}

Hamiltonian (\ref{H_KG}) is a conservative system and its total energy
$E\ge 0$ is preserved and serves as a control parameter of the
nonlinearity strength. Equations (\ref{H_EM}) become linear by
neglecting the nonlinear term $|u_{l}|^{\sigma} u_{l} $ or in the
limit $E\rightarrow 0$ (i.e.~$|u_{l}|\rightarrow 0$) where
$|u_{l}|^{\sigma} |u_{l}| \ll |u_{l}|  $. Then the ansatz
$u_{l} = A_{l} \exp(i\omega t)$ reduces them to the linear eigenvalue
problem
\begin{equation}
\omega^2 A_l=\left( \tilde{\epsilon}+\frac{2}{W} \right) A_l -\left(
\frac{1}{W} \right) \left( A_{l-1}+A_{l+1}\right).
\label{linear_eig}
\end{equation}
The normalized eigenvectors $A_{\nu,l}$ ($\sum_l A_{\nu,l}^2=1)$ are
the NMs of the system with the corresponding eigenvalues
$\lambda_{\nu}=\omega_{\nu}^2$. The width of the squared
eigenfrequency $\omega_{\nu}^2$ spectrum is $\Delta_{K}= 1+
\frac{4}{W}$ with $\omega_{\nu}^2 \in \left[ \frac{1}{2},\frac{3}{2} +
  \frac{4}{W}\right] $.

The asymptotic spatial decay of an eigenvector is given by $A_{\nu,l}
\sim {\rm e}^{-l/\xi(\lambda_{\nu})}$ where $\xi(\lambda_{\nu}) \leq
\xi(0) \approx 100/W^2$ is the localization length \cite{KM93}. The
spatial extend of a NM can be characterized by the average
localization volume $V\approx\sqrt{12 m_2}$, with $m_2$ being its
second moment \cite{D10}.  $V \approx 1$ for $W \gg 10$ , and $V
\approx 3.6\xi(0)$ for $W \leq 4$.  The average spacing of squared
eigenfrequencies of NMs within the range of a localization volume is
$d = \Delta_{K} / V \approx (W^2+4W)/360$ for $W \leq 4$ and $d \sim
\Delta_K$ for $W\gg 10$.

The squared frequency shift of a single-site oscillator induced by the
nonlinearity is
\begin{equation}
\delta_l=a_{\sigma} \left( \frac{E_l}{\tilde{\epsilon}_l}\right)^{\frac{\sigma}{2}}\;,\;
a_{\sigma}=
 \frac{2^{\frac{\sigma+4}{2}}}{\sqrt{\pi}(\sigma+2)} \frac{\Gamma
\left(\frac {\sigma+3}{2}\right) }{\Gamma \left(\frac {\sigma+2}{2}\right)}\;,
\label{dl}
\end{equation}
where $E_l$ is the energy of the oscillator (see Appendix
\ref{ap:shift}). For $\sigma=2$ in Eq.~(\ref{dl}) it follows $\delta_{l}
\approx (3 E_l)/(2 \tilde{\epsilon}_l)$ \cite{SKKF09}.

In our study $W=4$ and we follow the evolution of single site
excitations for long times and for $0< \sigma \leq 4$. Then we have
$\Delta_K=2$, $V\approx 20$ and $d\approx 0.1$.  We excite site $l_0$
by setting $p_l=\sqrt{2E}\delta_{l,l_0}$, $u_l=0$ for $t=0$ with
$\tilde{\epsilon}_{l_0}=1$. In our simulations we use symplectic
integration schemes of the SABA family of integrators \cite{SKKF09,
  LR01} with some particularities \cite{particular}.

In our computations the number of lattice sites $N$ and the
integration time step $\tau$ varied between $N=500$ to $N=3000$ and
$\tau=0.2$ to $\tau=0.05$, in order to exclude finite size effects in
the wave packet evolution and allow long integrations up to $10^{9}$
time units. In all our simulations the relative energy error was kept
smaller than $10^{-3}$.

Following the methodology of \cite{FKS09,SKKF09} we order the NMs in
space by increasing value of their center-of-norm coordinate
$X_{\nu}=\sum_l l A_{\nu,l}^2$. We consider normalized energy density
distributions $z_{\nu}\equiv E_{\nu}/\sum_{\mu} E_{\mu}$ with $E_{\nu}
= \dot{A}^2_{\nu}/2+\omega^2_{\nu}A_{\nu}^2/2$, where $A_{\nu}$ is the
amplitude of the $\nu$th NM and $\omega^2_{\nu}$ the corresponding
squared eigenfrequency. In our analysis we use the second moment $m_2=
\sum_{\nu} (\nu-\bar{\nu})^2 z_{\nu}$ ($\bar{{\nu}} = \sum_{\nu} \nu
z_{\nu}$), which quantifies the wave packet's degree of spreading, the
participation number $P=1 / \sum_{\nu} z_{\nu}^2$, which measures the
number of the strongest excited modes in $z_{\nu}$, and the
compactness index $\zeta=P^2/m_2$, which quantifies the sparseness of
a wave packet, since $\zeta$ decreases as the wave packet becomes more
sparse (see \cite{SKKF09} for more details).

\section {Different dynamical regimes}
\label{s:regimes}

Different dynamical regimes have been discussed in recent publications
\cite{KKFA08,PS08,FKS09,SKKF09}. In a recent paper one of us presented
a coherent interpretation of collected numerical data within a simple
framework which uses two averaged parameters of an initial wave packet
as essential control parameters for the dynamical evolution: the
average norm or energy density in a packet, and its typical size
\cite{F10}.  According to these results, a wave packet can be
selftrapped (see also \cite{KKFA08,SKKF09}) in a regime of strong
nonlinearity. This happens when nonlinear frequency shifts are larger
than the width of the spectrum of the linear equations.  If the
nonlinearity is weak enough to avoid selftrapping, the wave packet
will spread either in an intermediate regime of strong chaos, followed
by an asymptotic regime of weak chaos, or the strong chaos regime is
skipped, and spreading starts in the regime of weak chaos and stays
there. The outcome depends on the ratio of the nonlinear frequency
shift of the wave packet after spreading into the full localization
volume and the average spacing $d$.  Here we will adapt these
arguments to the study of single site excitations.  Then the three
possible regimes are:
\begin{eqnarray}
a_{\sigma} E^{\sigma/2} >  \Delta_K \;:\; {\rm selftrapping},
\label{selftrapping}
\\
a_{\sigma} \left(\frac{E}{V}\right)^{\sigma/2} > d \;:\; {\rm strong\; chaos},
\label{strongchaos}
\\
a_{\sigma} \left(\frac{E}{V}\right)^{\sigma/2} < d \;:\; {\rm weak\; chaos}.
\label{weakchaos}
\end{eqnarray}
Wave packets in the strong chaos regime spread faster than in the weak
chaos case (see also Sect.~\ref{sec:theory} below).

\begin{figure}
\includegraphics[angle=0,width=0.99\columnwidth]{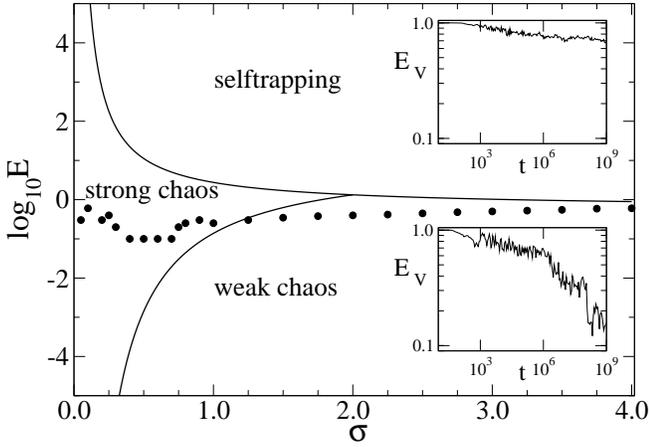}
\caption{The three different dynamical regimes for the gKG model, in
  the parameter space of the nonlinearity order $\sigma$ and the
  energy $E$ of a single nonlinear oscillator excitation. The points
  correspond to the particular numerical simulations presented in
  Sect.~\ref{sec:results}. Insets: The fraction $E_V$ of the wave
  packet's energy in a localization volume $V=20$ around the initially
  excited site $l_0$ versus time $t$ in log-log plots for $\sigma=2.5$
  and total energy $E=0.45$ (weak chaos regime, lower inset), $E=2.0$
  (selftrapped regime, upper inset). The disorder realization is the
  same as in Fig.~\ref{fig_3_regimes}. }
\label{fig_param_sp}
\end{figure}

\begin{figure}
\begin{tabular}{cc}
\includegraphics[scale=0.26]{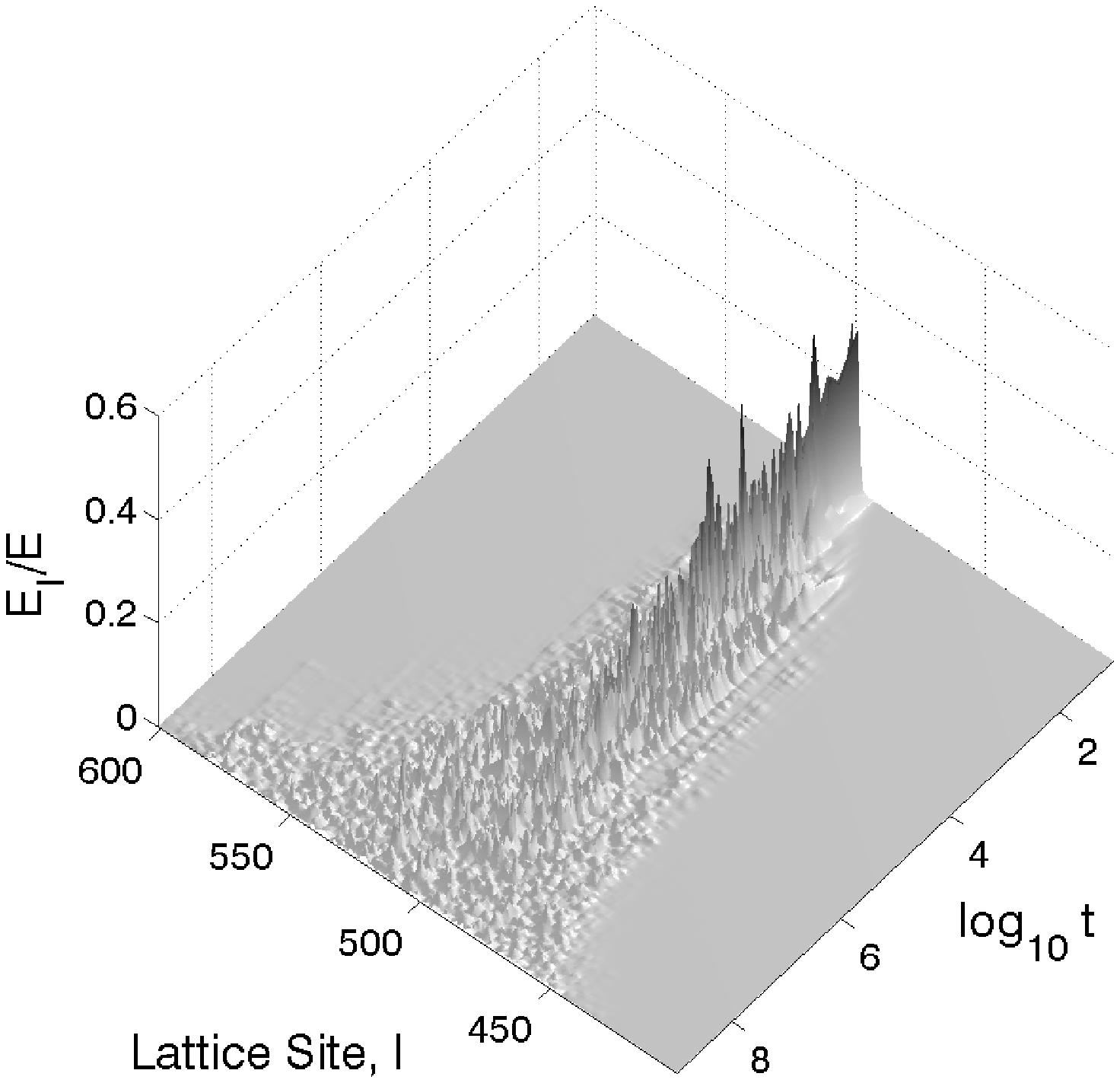} \vspace{0.6 cm} &
\includegraphics[scale=0.26]{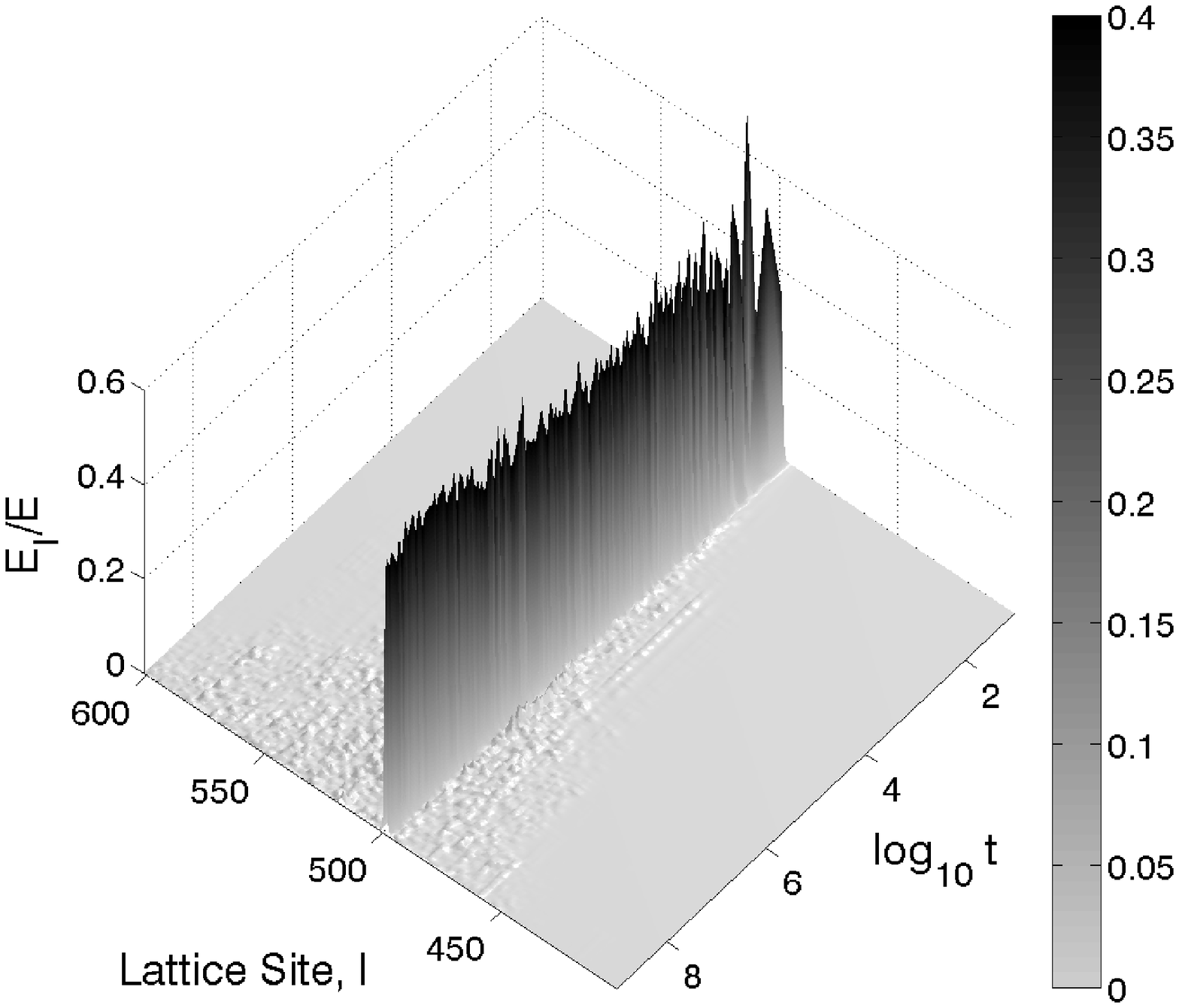} 
\end{tabular}
\caption{Time evolution of the normalized energy distribution $E_l/E$
  of wave packets in the neighborhood of the initially excited site
  $l_0=500$ for $\sigma=2.5$ and total energy $E=0.45$ (weak chaos
  regime, left plot), $E=2.0$ (selftrapped regime, right plot). Darker
  regions correspond to higher intensities. The disorder realization
  is the same as in Fig.~\ref{fig_3_regimes}.  }
\label{fig_3D}
\end{figure}

The location of the three different dynamical regimes in the parameter
space of the system's energy $E$ and the order $\sigma$ of the
nonlinearity is shown in Fig.~\ref{fig_param_sp}.  For $\sigma \geq 2$
the regime of strong chaos is absent (in the sense that if at all, it
will coexist with selftrapping).  Therefore, if not selftrapped, the
wave packet is expected to spread in the asymptotic regime of weak
chaos.  Following Anderson's definition of localization \cite{A58}, we
measure the fraction $E_V$ of the wave packet energy in a localization
volume $V=20$ around the initially excited site.  For a localized
state this fraction should asymptotically tend to a nonzero constant.
We find that in the weak chaos regime (lower inset of
Fig.~\ref{fig_param_sp}) the fraction continuously drops down in time,
indicating complete delocalization. Contrary, in the strong nonlinear
regime of selftrapping (upper inset of Fig.~\ref{fig_param_sp}) the
fraction appears to tend towards a nonzero constant. These behaviors
are also clearly seen in Fig.~\ref{fig_3D}.

For $\sigma \rightarrow 0$ the selftrapped regime is shifted
to very large energies.  At the same time for $\sigma < 2$ the regime
of strong chaos is widening its window with decreasing $\sigma$.  A
wave packet, if not selftrapped, may then spread in the intermediate
regime of strong chaos, and cross over to weak chaos at energy
densities which are the smaller the smaller $\sigma$ is.  Therefore
the crossover to the asymptotic regime of weak chaos may be pushed to
very large times, if the initial energy is fixed, and $\sigma$
lowered. For low enough initial energies the regime of strong chaos
can be again avoided. However low initial energies imply large time
scales which are needed to detect any type of spreading
\cite{FKS09,SKKF09}.  Since for $\sigma=0$ the equations become linear
again, Anderson localization is restored for all times.  Therefore, we
expect that for fixed initial energy and $\sigma \rightarrow 0$, the
characteristic time scales diverge as well.

\begin{figure}
\includegraphics[angle=0,width=0.99\columnwidth]{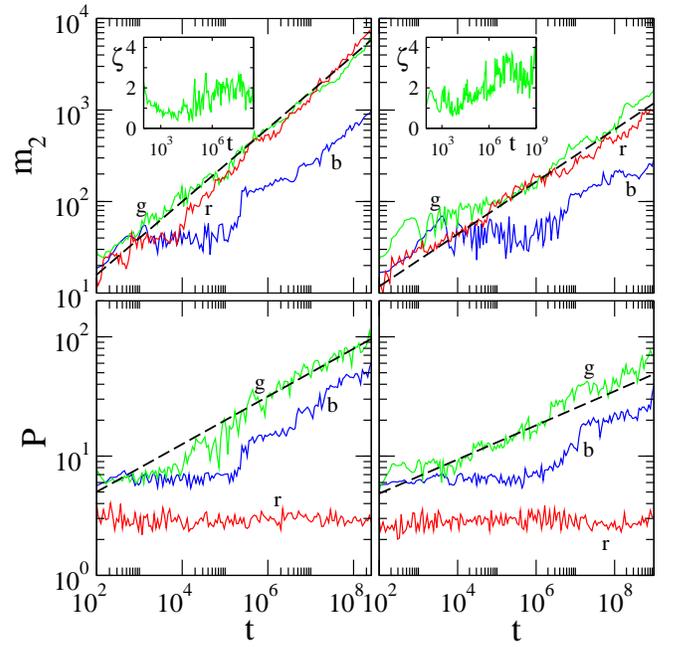}
\caption{(Color online) Different spreading behaviors.  $m_2$ and $P$
  versus time $t$ in log-log plots for different nonlinearity orders
  $\sigma$. Left plots: $\sigma=1.5$ and total energy
  $E=0.03,0.35,2.2$ [(b), blue; (g) green; (r) red]. Right plots:
  $\sigma=2.5$ and total energy $E=0.11,0.45,2.0$ [(b) blue; (g)
  green; (r) red].  The disorder realization is the same for both
  values of $\sigma$.  Straight lines guide the eye for exponents 2/5
  ($m_2$) and 1/5 ($P$) for $\sigma=1.5$, and 2/7 ($m_2$) and 1/7
  ($P$) for $\sigma=2.5$. Insets: the compactness index $\zeta$ as a
  function of time in linear-log plots for $E=0.35$ ($\sigma=1.5$) and
  $E=0.45$ ($\sigma=2.5$).}
\label{fig_3_regimes}
\end{figure}

Representative examples in the selftrapping and weak chaos regimes for
$\sigma=1.5$ and $\sigma=2.5$, are shown in Fig.~\ref{fig_3_regimes}.
In the regime of weak chaos wave packets initially evolve as in the
linear case: i.~e.~they show Anderson localization up to some time
$\tau_d$ and both $m_2$ and $P$ remain constant. For $t> \tau_d$ the
wave packet starts to grow with $m_2\sim t^{\alpha}$, $P\sim
t^{\alpha/2}$ (blue curves). The values $\alpha=2/5$ for $\sigma=1.5$
and $\alpha=2/7$ for $\sigma=2.5$, obtained by a theoretical
prediction for the weak chaos regime given in \cite{FKS09,F10} (see
also Sect.~\ref{sec:theory} below), describe quite well the numerical
data. Increasing the energy shortens the time $\tau_d$ (green
curves). The evolution of the compactness index $\zeta$ for these
cases is shown in the insets of Fig.~\ref{fig_3_regimes}. For both
values of $\sigma$ the compactness index eventually oscillates around
some constant non-zero value, implying that the wave packet does not
become more sparse when it spreads. In the regime of selftrapping a
large part of the wave packet remains localized, and therefore $P$ is
practically constant, while the rest spreads and the second moment
increases as $m_2\sim t^{\alpha}$ (red curves). The time evolution of
the energy fraction $E_V$ contained in a localization volume $V=20$
around the initially excited site for $\sigma=2.5$ in the cases of
immediate subdiffusion and of selftrapping is shown in the insets of
Fig.~\ref{fig_param_sp}.

\section {Spreading of wave packets}
\label{sec:spreading}

\subsection {Theoretical predictions}
\label{sec:theory}

In the regime of weak chaos, only a fraction of modes in the packet
resonantly interact and facilitate randomization of phases
\cite{FKS09,SKKF09,F10}.  The second moment of the wave packet
increases in time and follows a power law of the form $m_2\sim
t^{\alpha}$. In these cases, the participation number follows the law
$P\sim t^{\alpha/2}$. The dependence of the exponent $\alpha$ on the
order $\sigma$ of the nonlinearity was predicted to be
\cite{FKS09,F10}
\begin{equation}
\alpha=\frac{1}{1+\sigma}\;.
\label{alpha_normal}
\end{equation}
The derivation is based on the result that the probability
$\mathcal{P}$ of resonance for a given packet mode depends on the
energy density $\epsilon$ in the packet.  For $a_{\sigma}
\epsilon^{\sigma/2} > d$ (strong chaos) it follows $\mathcal{P}
\approx 1$, while for $a_{\sigma} \epsilon^{\sigma/2} \ll d$ (weak
chaos) we get $\mathcal{P} \approx \frac{a_{\sigma}
  \epsilon^{\sigma/2}}{d}$ \cite{F10}.  The diffusion rate is
conjectured to be $D \sim \epsilon^{\sigma}
(\mathcal{P}(\epsilon))^2$, which in the case of weak chaos, leads to
(\ref{alpha_normal}) together with $m_2 \sim 1/\epsilon$.

In the regime of strong chaos we get $\mathcal{P}=1$ and the exponent
$\alpha$ of $m_2 \sim t^{\alpha}$ is given by \cite{FKS09,F10}
\begin{equation}
\alpha=\frac{2}{2+\sigma}.
\label{alpha_deph}
\end{equation}
Since the energy density in the packet is decreasing with increasing
time, the condition for strong chaos will be eventually violated, and
the spreading will cross over into the regime of weak chaos
\cite{F10}. The crossover duration can be very large, and complicates
the fitting analysis of numerical data.

If the phases of normal modes are randomized by an explicit routine
during the evolution of the system, then the regime of strong chaos is
enforced irrespectively on whether the energy density satisfies the
criterion of strong or weak chaos. In that case the spreading should
again follow the law (\ref{alpha_deph}) but now for all times
\cite{FKS09}.

The above predictions are expected to hold if a coherent transfer of
energy can be neglected, and only incoherent diffusive transfer is
relevant.

In previous studies with $\sigma=2$ and single site excitations,
spreading should start in the regime of weak chaos.  Corresponding
fits of the exponent yield $\alpha=0.33\pm 0.05$ for the KG model
\cite{FKS09,SKKF09}, in agreement with Eq.~(\ref{alpha_normal}).
Mulansky computed spreading exponents for the gDNLS model with single
site excitations and $\sigma =1,2,4,6$ \cite{M09}. Since for
$\sigma=2,4,6$ strong chaos is avoided, the fitting of the dependence
$m_2(t)$ with a single power law is reasonable.  The corresponding
fitted exponents $0.31\pm0.04$ ($\sigma=2$), $0.18 \pm 0.04$
($\sigma=4$) and $0.14 \pm 0.05$ ($\sigma=6$) agree well with the
predicted weak chaos result $1/3,1/5,1/7$ from (\ref{alpha_normal}).
For $\sigma=1$ the initial condition has been launched in the regime
of strong chaos. A single power law fit will therefore not be
reasonable. Since the outcome is a mixture of first strong and later
possibly weak chaos, the fitted exponent should be a number which is
located between the two theoretical values $1/2$ and $2/3$.  Indeed,
Mulansky reports a number $0.56\pm 0.04$.  Veksler et al.~\cite{VKF09}
considered short time evolutions of single site excitations (up to
$t=10^3$) for gDNLS models.  While the time window may happen to be
too short for conclusive results, the observed increase of fitted
exponents with increasing $\beta$ for $\sigma < 2$ is possibly also
influenced by the transition from weak to strong chaos.

\subsection {Numerical results}
\label{sec:results}

In our study we provide numerical evidence for the validity of both
predictions (\ref{alpha_normal}) and (\ref{alpha_deph}), by performing
extensive simulations of the gKG model for various values of $\sigma$
and for energies $E$ away from the selftrapping regime. We consider
not only integer but also noninteger values of $\sigma$.

The used energies $E$ are not too small in order to avoid long delays
on the onset of spreading, and their values cross the boundary between
the weak and strong chaos regimes as $\sigma$ decreases
(Fig.~\ref{fig_param_sp}). The transition from weak to strong chaos is
not an abrupt one. One should keep in mind that the curves plotted in
Fig.~\ref{fig_param_sp} should be considered as rough indicators of
the borders between different regimes, since they are influenced by
the characteristics of each particular disorder realization.

\subsubsection {Computational techniques}
\label{sec:techniques}

For several values of the nonlinearity order $\sigma$ we compute the
evolution of single site excitations up to a large final time
$t_{fin}$ for an energy value which excludes selftrapping, and for 20
different disorder realizations. For each case we verify that the time
evolution of $m_2$ and $P$ indicate that the wave packet is not
selftrapped.  Although we also compute $P$ we chose to analyze and
present results only for $m_2$, because it grows faster than $P$
allowing a more accurate determination of the exponent $\alpha$.
Typically $t_{fin}$ ranges from $t_{fin}=10^8$ for small $\sigma$ to
$t_{fin}=10^9$ for larger $\sigma$ for which slower spreading is
observed.  For each realization the $m_2(t)$ is fitted by a
$t^{\alpha}$ law, the value of the exponent $\alpha$ is determined and
the average value $\langle \alpha \rangle$ over the 20 realizations is
computed. In practice, we perform a linear fit of the $\log_{10}
m_2(t)$, $\log_{10} t$ values, instead of a nonlinear fit of the
actual $m_2(t)$, $t$ values. The data used for the fitting lie in the
time window $[\log_{10} t_{ini},\log_{10} t_{fin}]$ whose lower end
varies from $\log_{10} t_{ini}=1$ up to $\log_{10} t_{ini}=\log_{10}
t_{fin}-2$, in order to guarantee that even the smallest window
contains enough data points (for at least two orders of magnitude of
$t$) allowing a reliable evaluation of $\alpha$.

\begin{figure}
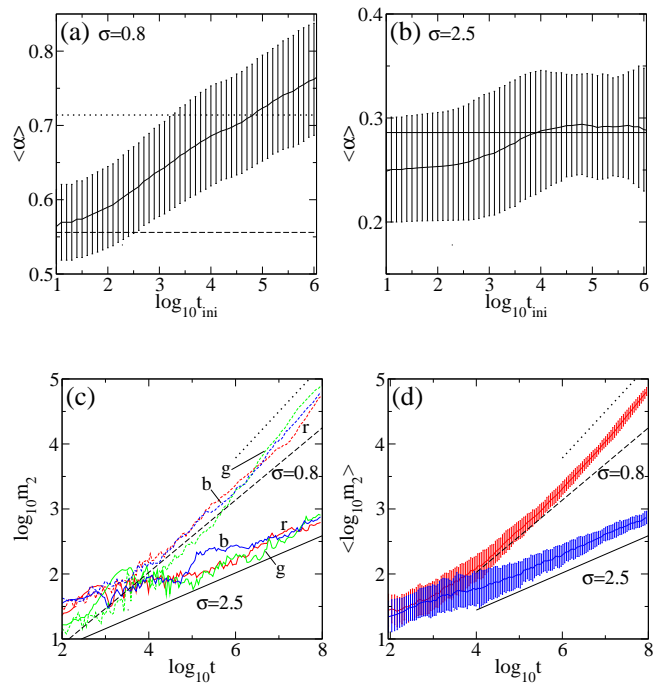

\begin{tabular}{cc}
\includegraphics[scale=0.229]{f_4_a.eps} \vspace{0.6 cm} &
\includegraphics[scale=0.229]{f_4_b.eps}  \\ 
\includegraphics[scale=0.229]{f_4_c.eps} &
\includegraphics[scale=0.229]{f_4_d.eps}
\end{tabular}
\caption{(Color online) The numerically obtained mean value $\langle
  \alpha \rangle $ of the exponent $\alpha$ ($m_2 \sim t^{\alpha}$)
  over 20 disorder realizations as a function of the initial time
  $t_{ini}$ of the fitting window having as final time $t_{fin}=10^8$
  in linear-log plots for (a) $\sigma=0.8$, $E=0.25$ and (b)
  $\sigma=2.5$, $E=0.45$. (c) $\log_{10} m_2$ versus $\log_{10} t$ for
  three different disorder realizations (denoted by (b) blue, (g)
  green, and (r) red) for $\sigma=0.8$ (dashed upper curves) and
  $\sigma=2.5$ (solid lower curves). (d) The mean value $\langle
  \log_{10} m_2 \rangle $ of $\log_{10} m_2$ (with error bars) over 20
  disorder realizations versus $\log_{10} t$ for $\sigma=0.8$ (upper
  curve) and $\sigma=2.5$ (lower curve). In all panels straight lines
  correspond to theoretical predictions for the values of
  $\alpha$. For $\sigma=0.8$ the weak chaos exponent $\alpha=5/9$ from
  Eq.~(\ref{alpha_normal}) (dashed lines) and the strong chaos
  exponent $\alpha=5/7$ from Eq.~(\ref{alpha_deph}) (dotted lines) are
  plotted, while for $\sigma=2.5$ the weak chaos exponent $\alpha=2/7$
  (solid lines) is shown. }
\label{fig_compute_slopes}
\end{figure}

If the time evolution of $m_2(t)$ is well approximated by a
$t^{\alpha}$ law the averaged value $\langle \alpha \rangle$ should
not depend on the initial time $t_{ini}$ of the time window. In other
words, if $\langle \alpha \rangle $ eventually becomes a constant
function of $\log_{10} t_{ini}$ for large enough values of $t_{ini}$
then we consider that $t^{\langle \alpha \rangle}$ satisfactorily
describes the asymptotic ($t\rightarrow \infty$) behavior of
$m_2(t)$. If on the other hand, $\langle \alpha \rangle $ does not
tend to become constant as $t_{ini}$ increases, the available
numerical data cannot be represented reliably by a $t^{\alpha}$ fit
and no exponent can be determined. Such an example is seen in
Fig.~\ref{fig_compute_slopes}(a) where we see that for $\sigma=0.8 $
$\langle \alpha \rangle $ increases monotonically as $t_{ini}$
increases.  The horizontal dashed line denotes the theoretically
predicted weak chaos exponent from Eq.~(\ref{alpha_normal})
$\alpha=5/9=0.556$.  At the largest values of $t_{ini}$ the exponent
is close to the strong chaos result $\alpha=5/7=0.714$ from
Eq.~(\ref{alpha_deph}) (horizontal dotted line).

For $\sigma=2.5$ (Fig.~\ref{fig_compute_slopes}(b)) the values of
$\langle \alpha \rangle $, after some initial transient time, seem to
saturate to a constant value implying that a $t^{\alpha}$ law well
approximates the evolution of $m_2$. Actually, $\langle \alpha \rangle
$ eventually attains values close to the theoretically predicted weak
chaos exponent $\alpha=2/7=0.286 $, denoted by a horizontal line in
Fig.~\ref{fig_compute_slopes}(b).

A significant dynamical quantity is the minimum time $\tau^*$ after
which $\langle \alpha \rangle $ becomes practically constant, since it
characterizes the onset of the validity of the asymptotic
approximation $m_2 \sim t^{\alpha}$. The value of $\tau^*$ is
estimated from the numerically evaluated function $\langle \alpha
\rangle =f(\log_{10} t_{ini})$ as the minimum time such that for all
$t_{ini}>\tau^*$ the local (numerically estimated) rate $d(\langle
\alpha \rangle)/d(\log_{10} t_{ini})$ guarantees a less than 5\%
change of $\langle \alpha \rangle $ at $t_{fin}$ with respect to its
current value at $\tau^*$. This definition has a degree of
arbitrariness but it manages to capture for all our simulations the
time after which a $t^{\alpha}$ fit seems to approximate quite
accurately the values of $ m_2(t)$. For example for $\sigma=2.5$
(Fig.~\ref{fig_compute_slopes}(b)), we find $\log_{10} \tau^*=4$, and
the corresponding exponent value $\langle \alpha\rangle =0.288\pm
0.058$ is very close to the theoretically predicted value
$\alpha=0.286$ from Eq.~(\ref{alpha_normal}).

The cases of $\sigma=2.5$ and $\sigma=0.8$ shown in
Fig.~\ref{fig_compute_slopes} are two typical examples where an
exponent $\alpha$ for $m_2(t)\sim t^{\alpha}$ can or cannot be defined
respectively. The time evolution of $m_2$ for three particular
disorder realizations for $\sigma=2.5$ and $\sigma=0.8$ is plotted in
Fig.~\ref{fig_compute_slopes}(c). For $\sigma=2.5$ the three curves
tend to increase according to the theoretical prediction given in
Eq.~(\ref{alpha_normal}) (solid line). On the other hand, for
$\sigma=0.8$ $m_2$ seems to increase with an increasing rate, not
showing a tendency to approach any power law, at least up to the final
integration time $t_{fin}=10^8$. In this case, $m_2$ deviates from the
weak chaos theoretical power law $t^{5/9}$ prediction given in
Eq.~(\ref{alpha_normal}) (dashed line). Although a reliable numerical
estimation of a constant $\langle \alpha \rangle $ was not possible
from our simulations, the values of $m_2$ for large values of $t$ seem
to be close to the strong chaos prediction $m_2 \sim t^{5/7}$ given
from Eq.~(\ref{alpha_deph}) (dotted line). The deviation from the
power law predictions for $\sigma=0.8$ can also be clearly seen by
plotting the mean value $\langle \log_{10} m_2 \rangle $ of $\log_{10}
m_2(t)$ over the 20 realizations as a function of time
(Fig.~\ref{fig_compute_slopes}(d)). Similar results for $\sigma=2.5$
show that the numerical results are in good agreement with the weak
chaos prediction $m_2 \sim t^{2/7}$.

\subsubsection {Subdiffusive spreading}
\label{sec:sub}

We performed extensive simulations for 25 different values of the
nonlinearity order $\sigma$ in the interval $0<\sigma \leq 4$.  For
each $\sigma$, we followed the evolution of single site excitations
for 20 different disorder realizations by considering an energy value
away from the selftrapping regime, which allows the immediate
subdiffusion of the wave packet. For each case we tried to determine
the exponent $\alpha$ ($m_2 \sim t^{\alpha}$) and the time $\tau^*$ by
the above-described procedure. Our criterion for determining $\alpha$
and $\tau^*$ was satisfied for $\sigma=0.05$ and $1.25 \leq \sigma
\leq 4$. For all other tested values of $\sigma$ ($0.1 \leq \sigma\leq
1$) a reliable value for $\alpha$ was not obtained, since $m_2$
exhibited behaviors similar to the one seen for $\sigma=0.8$
(Fig.~\ref{fig_compute_slopes}).  It is possible that for these cases
integration up to larger final times might allow $m_2$ to attain its
asymptotic power law behavior, and permit the estimation of $\alpha$,
but the extremely long CPU times needed for such simulations do not
make them easily feasible. As we can see from Fig.~\ref{fig_param_sp},
all these cases belong to the strong chaos regime. The fact that
exactly for $\sigma < 2$ the intermediate (and possibly rather long
lasting) regime of strong chaos may set in, is a possible explanation
for the observed difficulties.

\begin{figure}
\includegraphics[angle=0,width=0.99\columnwidth]{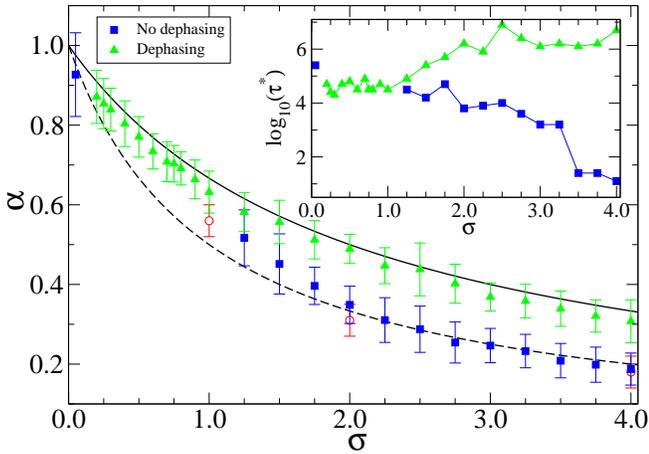}
\caption{(Color online) Exponent $\alpha$ ($m_2 \sim t^{\alpha}$)
  versus the nonlinearity order $\sigma$ for plain integration without
  dephasing (filled squares) and for integration with dephasing of NMs
  (filled triangles). Results without dephasing obtained in \cite{M09}
  are plotted with empty circle symbols. The theoretically predicted
  functions $\alpha=1/(1+\sigma)$ (weak chaos) and
  $\alpha=2/(2+\sigma)$ (strong chaos) are plotted by dashed and solid
  lines respectively. Inset: The logarithm of the minimum time
  $\tau^*$ for which the evolution of $m_2$ can be numerically fitted
  by a function of the form $t^{\alpha}$ versus $\sigma$ for
  integration with (filled triangles) and without (filled squares)
  dephasing.}
\label{fig_alpha}
\end{figure}

The computed exponents $\alpha$ for different values of $\sigma$ are
plotted in Fig.~\ref{fig_alpha} (filled squares). The values of
$\alpha$ obtained in \cite{M09} for gDNLS are also plotted (empty
circles). These values are very close to our results for $\sigma=2$
and $\sigma=4$, while for $\sigma=1$ a reliable estimate of $\alpha$
was not obtained from our simulations. In Fig.~\ref{fig_alpha} the
theoretically predicted law (\ref{alpha_normal}) is plotted by a
dashed line and all computed exponents $\alpha$ are in good agreement
with it. The exponents for $\sigma=1.75$, $\sigma=1.5$ and
$\sigma=1.25$ slightly deviate from this law, possibly due to the
influence of the strong chaos regime which exists for $\sigma <2$. The
time $\tau^*$ after which $m_2$ is well approximated by $t^{\alpha}$
(inset of Fig.~\ref{fig_alpha}), increases as $\sigma$ approaches
zero, implying that integration for longer times might be needed in
order to estimate $\alpha$ for $0.1 \leq \sigma\leq 1$.

\begin{figure}[t]
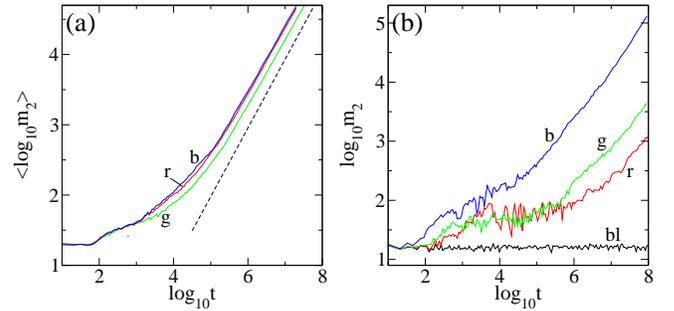

\begin{tabular}{cc}
\includegraphics[scale=0.229]{f_6_a.eps} &
\includegraphics[scale=0.229]{f_6_b.eps} 
\end{tabular}
\caption{(Color online) (a) mean value $\langle \log_{10} m_2 \rangle$
  of $\log_{10} m_2$ over 20 realizations versus $\log_{10} t$ for
  $\sigma=0.05$ and energy $E=0.3, 5000, 50000$ [(g) green; (r) red;
  (b) blue]. Straight line guide the eye for the theoretically
  predicted strong chaos exponent $\alpha=0.976$. (b) $\log_{10} m_2$
  versus $\log_{10} t$ for the same disorder realization with $E=0.03$
  and $\sigma=0.00001, 0.005, 0.01, 0.05$ [(bl) black; (r) red; (g)
  green; (b) blue].  }
\label{fig_s_2_05}
\end{figure}

From Fig.~\ref{fig_param_sp} we expect that for small values of
$\sigma$ the range of energies $E$ for which we observe immediate
subdiffusive spreading in the regime of strong chaos should
increase. As an example we consider the case of $\sigma=0.05$, for
which the exponent $\alpha$ was obtained for $E=0.3$
(Fig.~\ref{fig_alpha}). In Fig.~\ref{fig_s_2_05}(a) the evolution of
$\langle \log_{10} m_2 \rangle $ (average value over 20 realizations)
is plotted for $E=0.3$ (green curve), $E=5000.0$ (red curve) and
$E=50000.0$ (blue curve). For all energies $\langle \log_{10}
m_2\rangle $ increases in a similar way following a power law, after
some transient initial time interval, which is well approximated by
the theoretically predicted function (\ref{alpha_deph}) $t^{0.976}$
(dashed line). It is evident that the exponent $\alpha$, which for
$E=0.3$ was estimated to be $\alpha=0.93\pm 0.11$ at $\log_{10}
\tau^*=5.4$, does not depend on the energy value. Note that the error
bar is too large here to discriminate between the strong chaos
$\alpha=0.976$ and weak chaos $\alpha=0.952$ predictions.

For very small values of $\sigma$ the dynamics should approach the
behavior of the linear system ($\sigma=0$), i.~e.~the wave packet
should be localized. This tendency is seen in Fig.~\ref{fig_s_2_05}(b)
where $\log_{10} m_2(t)$ is plotted for a disorder realization with
$E=0.3$ in the cases of $\sigma= 0.05$, $0.01$, $0.005$, $0.00001$
(curves from top to bottom). It is evident that the time $\tau^*$
increases as $\sigma\rightarrow0$. In particular, for $\sigma=0.00001$
this time was not reached until the end of the simulation at $t=10^8$.

\begin{figure}
\includegraphics[angle=0,width=0.99\columnwidth]{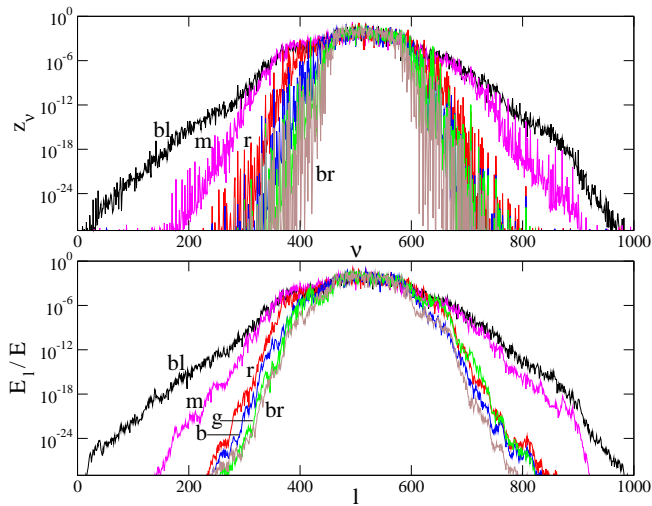}
\caption{(Color online) Normalized energy distributions in NM (upper
  plot) and real (lower plot) space for $\sigma=0.05, 0.2, 0.8, 1.25,
  2.0, 3.0$ [(bl) black; (m) magenta; (r) red; (b) blue; (g) green;
  (br) brown] at times $t=3.6\times 10^5, 1.3 \times 10^5, 2.5 \times
  10^5, 1.4 \times 10^6, 3 \times 10^7, 10^9$ respectively. The second
  moment of each distribution is $m_2\approx10^3$. In the upper plot
  the distributions for $\sigma=1.25, 2.0$ are not clearly visible as
  they are overlapped by the distribution for $\sigma=3.0$.}
\label{profiles}
\end{figure}

The order of nonlinearity $\sigma$ influences not only the spreading
rate of wave packets, but also the morphology of their profiles. In
Fig.~\ref{profiles} we plot the normalized energy distributions of
initial single site excitations, for different $\sigma$ values in NM
(upper plot) and real (lower plot) space. Starting from the outer,
most extended wave packet we plot distributions for $\sigma=0.05$
(black curves), $\sigma=0.2$ (magenta curves), $\sigma=0.8$ (red
curves), $\sigma=1.25$ (blue curves), $\sigma=2$ (green curves) and
$\sigma=3$ (brown curves). All wave packets were considered for the
same disorder realization but at different times of their evolution
when they have the same value of second moment $m_2\approx10^3$. These
times are $t=3.6\times 10^5$ for $\sigma=0.05$, $t=1.3\times 10^5$ for
$\sigma=0.2$, $t=2.5\times 10^5$ for $\sigma=0.8$, $t=1.4\times 10^6$
for $\sigma=1.25$, $t=3\times 10^7$ for $\sigma=2$ and $t=10^9$ for
$\sigma=3$ and increase for $\sigma \geq 0.2$ since the spreading
becomes slower for larger $\sigma$. As we have seen in
Fig.~\ref{fig_s_2_05}, when $\sigma\rightarrow0$ wave packets remain
localized for very large time intervals before they start to
spread. This is why for $\sigma=0.05$ the second moment becomes
$m_2\approx10^3$ at a larger time than in cases with $\sigma=0.2$ and
$\sigma=0.8$.

From the results of Fig.~\ref{profiles} we see that for large enough
values of $\sigma$ ($0.8 \leq \sigma \leq 3$), the distributions on a
logarithmic scale have a chapeau-like shape consisting of a highly
excited central part and exponential tails having practically the same
slope. Contrarily, the distributions for $\sigma =0.2$ and
$\sigma=0.05$ become more extended having different slopes in the
tails.

A characteristic of the NM space distributions in Fig.\ref{profiles}
for $\sigma \geq 0.8 $ is that they exhibit very large value
fluctuations (up to 10-15 orders of magnitude) in their tails,
contrarily to the corresponding distributions in real space.  Tail NMs
are driven by the core of the wave packet, and may also interact with
neighboring tail NMs. The presence of large tail amplitude
fluctuations signals that neighboring tail NMs do not interact
significantly (otherwise we would expect a tendency towards
equipartition).  Tail NMs are then excited only by the core; the
further away they are, the weaker the excitation. But within a small
tail volume, NMs with larger localization length will be more strongly
excited than those with smaller localization length, hence the large
observed fluctuations, which on a logarithmic scale are of the order
of the relative variation of the localization length.  Therefore
Anderson localization is preserved in the tails of the distributions
over very long times (essentially until the given tail volume becomes
a part of the core).  But the NM space distributions for $\sigma=0.05$
and $\sigma =0.2$ exhibit less fluctuations in their tail values with
respect to the other distributions in the upper plot of
Fig.~\ref{profiles}, implying that tail NMs are now interacting with
each other on comparatively short time scales and reach a visible
level of local equipartition. Therefore we observe for these cases a
destruction of Anderson localization even in the tails of the
spreading wave packets.

\subsubsection {Dephasing}
\label{sec:dephasing}

The assumption that all NMs in a wave packet are chaotic leads to a
power law increase of $m_2$ whose exponent $\alpha$ is given by
Eq.~(\ref{alpha_deph}). In order to check the validity of this
prediction we enhanced the wave packet chaoticity by a periodic
dephasing of its NMs. Every 100 time units on average 50\% of the NMs
were randomly chosen and the signs of their momenta were changed. In
this way a faster spreading of the wave packet, with respect to its
normal evolution, was achieved.  We also note that $\sigma$
has a similar effect on the shape of wave packets as in the case of
normal evolution without dephasing, because energy distributions in NM
and in real space for the same $m_2$ values and different $\sigma$
values have similar profiles to the ones shown in Fig.~\ref{profiles}.

Performing a similar numerical analysis (as in the case of normal wave
packet evolution) we computed the exponent $\alpha$ of $m_2\sim
t^{\alpha}$ and the time $\tau^*$ for several values of the
nonlinearity order $\sigma$. The obtained values are plotted in
Fig.~\ref{fig_alpha} by filled triangles. The numerically computed
exponents are in good agreement with the theoretical prediction of
Eq.~(\ref{alpha_deph}) (solid line in Fig.~\ref{fig_alpha}). In the
case of normal wave packet evolution not all NMs in the packet are
chaotic, because the exponents presented by filled squares in
Fig.~\ref{fig_alpha} are always smaller that the prediction of
Eq.~(\ref{alpha_deph}). The dephasing procedure increases drastically
the chaotic nature of the dynamics since the corresponding exponents
$\alpha$ are quite close, but somewhat smaller, than the predicted
values given by Eq.~(\ref{alpha_deph}) (strong chaos regime).

From the results of Fig.~\ref{fig_alpha} we see that exponents
$\alpha$ were determined for a larger value interval of $\sigma$ ($0.2
\leq \sigma \leq 4$) with respect to the normal evolution case. Only
for $\sigma=0.05$ and $\sigma=0.1$ we were not able to estimate an
exponent $\alpha$ from the performed numerical simulations.

In addition, time $\tau^*$ (inset of Fig.~\ref{fig_alpha}) has always
larger values with respect to the ones obtained for the normal
evolution of wave packets, especially for large values of
$\sigma$. This means that although dephasing increases the chaoticity
of the wave packet, a considerably large amount of time is needed for
the evolution to be characterized by $m_2\sim t^{\alpha}$.

\section {Summary and discussion}
\label{sec:sum}

We performed extensive numerical simulations of the evolution of
single site excitations in the gKG model (\ref{H_KG}) for different
values of the nonlinearity order $\sigma$. According to the analytical
treatment presented in \cite{F10}, in this case a wave packet could
either: a) be selftrapped for large enough values of the nonlinearity,
i.~e.~the total energy $E$ of the gKG system, or b) spread
subdiffusively for smaller values of $E$. Particularly for energy
values not in the selftrapping regime, the single site excitation
belongs either to the weak or the strong chaos regime \cite{F10}.

In the weak chaos regime the wave packet spreads subdiffusively and
its second moment $m_2$ grows as $m_2 \sim t^{\alpha}$. The expected
dependence of $\alpha$ on $\sigma$ in this case is given by
Eq.~(\ref{alpha_normal}). The detrapping time $\tau_d$ after which
spreading stars, increases as $E$ decreases because the system is
closer to a linear model where no spreading is observed due to
Anderson localization. In order to be able to observe spreading for
large time intervals we avoided very small energy values.

If the wave packet is launched in the strong chaos regime its
subdiffusive spreading is initially characterized by an exponent
$\alpha$ given from Eq.~(\ref{alpha_deph}), which is expected to
eventually cross over to its asymptotic ($t\rightarrow \infty$) value
given by Eq.~(\ref{alpha_normal}). The time at which this crossover
starts (as well as its duration) could become very large, limiting our
ability to observe it numerically.

According to the estimations of \cite{F10}, if the single site
excitation is not selftrapped then it belongs to the weak chaos regime
for $\sigma \geq 2$. For $\sigma <2$ the existence of the strong chaos
regime is possible, but not for very small energy values where the
dynamics is again in the weak chaos regime (see
Fig.~\ref{fig_param_sp}).

We performed numerical simulations for various values of $\sigma$ and
for $E$ belonging both to the weak and the strong chaos regimes,
having care to avoid very small energy values where we might encounter
very large detrapping times (Fig.~\ref{fig_param_sp}). The curves in
Fig.~\ref{fig_param_sp} do not define exactly the separation between
different regimes, instead they should be considered as indications of
the location of border regions between them. The energy values used
here cross the boundary between the weak and strong chaos regimes
around the interval $1 \lesssim \sigma \lesssim 2$.

Our results verify the validity of Eq.~(\ref{alpha_normal}) for the
weak chaos regime. In particular, the numerically computed exponents
$\alpha$ are in very good agreement with the theoretical prediction of
Eq.~(\ref{alpha_normal}) for $\sigma \geq 2$, while they exhibit an
increasing deviation from this prediction for $\sigma=1.75$,
$\sigma=1.5$ and $\sigma =1.25$ respectively
(Fig.~\ref{fig_alpha}). For $\sigma \leq 1$ $m_2$ grows faster than
the weak chaos estimation (\ref{alpha_normal}), but up to the final
times that we performed computations we were not able to reliably fit
its evolution with a power law and compute an exponent
$\alpha$. Nevertheless, this problem, as well as the deviation of the
computed exponents from the theoretical weak chaos prediction
(\ref{alpha_normal}) for $1.25 \leq \sigma \leq 1.75$, clearly
indicate that in our simulations the boundary between the weak and
strong chaos regimes lies in the interval $1 \lesssim \sigma \lesssim
2$, being in agreement with the theoretical predictions of
Fig.~\ref{fig_param_sp}.

Regarding the simulations for $\sigma \leq 1$ we note that maybe
longer integrations could allow for a reliable estimation of
$\alpha$. This is a very hard computational task as it requires large
CPU times. Since simulations with $\sigma$ values around $\sigma=1$
are located close to the borders of different regimes, it is possible
that for different disorder realizations the wave packet's evolution
is a mixture of different dynamical behaviors, not allowing the
statistical analysis to clearly determine $\alpha$. In addition, for
small values of $\sigma$ we observe differences in the wave packet
dynamics, which could affect the determination of $\alpha$, since
Anderson localization is destroyed also in the tails of the wave
packets (Fig~\ref{profiles}).

In order to enforce the wave packet to evolve continuously in the
strong chaos regime we enhanced its chaoticity by repeatedly
performing a dephasing of its NMs. When dephasing was applied, the
computed exponents $\alpha$ were in good agreement with the strong
chaos theoretical prediction of Eq.~(\ref{alpha_deph}) for almost all
tested values of $\sigma$ (Fig.~\ref{fig_alpha}).

Predictions (\ref{alpha_normal}) and (\ref{alpha_deph}) were derived
in Refs. \cite{FKS09,F10} for the gDNLS model (\ref{gDNLS}) and for
integer values of $\sigma$. Thus, as a final remark we note that our
results also establish the generality of these predictions since they
proved to be valid for a different dynamical system (the gKG model
(\ref{H_KG})) and for noninteger values of $\sigma$.

\begin{acknowledgments}
  We thank J.~D.~Bodyfelt, S.~Fishman, D.~O.~Krimer, Y.~Krivolapov,
  T.~Lapteva, N.~Li, M.~Mulansky, A.~Ponno and H.~Veksler for useful
  discussions.
\end{acknowledgments}

\appendix

\section{Frequency shift of nonlinear oscillators}
\label{ap:shift}

Let us consider a nonlinear oscillator described by the Hamiltonian
function
\begin{equation}
H=\frac{p^2}{2}+ \frac{\omega^2 x^2}{2}+ \frac{\beta}{\sigma+2} |x|^{\sigma+2},
\label{eq:osc}
\end{equation} 
with $\beta, \sigma \ge 0$. For a given value of the energy $E\ge
0$ the oscillator's period $T$ is
\begin{equation}
T=4\int_0^{\hat{x}} \frac{dx}{\sqrt{2E-\omega^2 x^2-\frac{2\beta}{\sigma+2} x^{\sigma+2}}} ,
\label{eq:T1}
\end{equation} 
where $\hat{x}$ is the positive root of equation $\omega^2
x^2+\frac{2\beta}{\sigma+2} |x|^{\sigma+2} =2E$. The change of variable $x=\sqrt{2E} v \hat{y}(\eta)/\omega$, with
\begin{equation}
\eta=\frac{2\beta}{\sigma+2} \frac{(2E)^{\sigma/2}}{\omega^{\sigma+2}}, 
\label{eq:eta}
\end{equation} 
and $\hat{y}(\eta)$ being the positive number satisfying
\begin{equation}
\hat{y}(\eta)^2+\eta |\hat{y}(\eta)|^{\sigma+2} =1,
\label{eq:root2}
\end{equation} 
transforms Eq.~(\ref{eq:T1}) to
\begin{equation}
  T(\eta)=\frac{4 \hat{y}(\eta)}{\omega} \int_0^{1} \frac{dv}{\sqrt{1-\hat{y}(\eta)^2 v^2-\eta \hat{y}(\eta)^{\sigma+2} v^{\sigma+2}}}.
\label{eq:T2}
\end{equation}

For $\beta=0$ we get $\eta=0$, $\hat{y}=1$ and Eq.~(\ref{eq:osc})
corresponds to an harmonic oscillator with period $T(0)=2\pi/ \omega$
and frequency $\Omega(0)=\omega$. In order to estimate the change of
the squared frequency $\delta \Omega^2=\Omega^2(\eta)-\Omega^2(0)$, we
note that the period of the oscillator for $\eta \ge 0$ is
$T(\eta)\approx T(0)+\eta T'(0)$, at a first order approximation in
$\eta$, with `$\,\,' \,\,$' denoting derivative with respect to
$\eta$. Differentiating Eq.~(\ref{eq:root2}) we find
$\hat{y}'(0)=-1/2$. Using this result, and differentiating
Eq.~(\ref{eq:T2}) according to the Leibniz integral rule we get
\begin{equation}
T'(0)=-\frac{2 \pi}{\omega}\left[   \frac{1}{2} +\frac{1}{\pi} \int_0^1
\frac{v^2-v^{\sigma+2}}{(1-v^2)^{3/2}} dv     \right]. 
\label{eq:T_der}
\end{equation} 
Then,  frequency $\Omega(\eta)$ is given by
\begin{equation}
  \Omega(\eta)=\frac{2\pi}{T(\eta)} \approx \omega \left[ 1+ \eta \left(   \frac{1}{2} +\frac{1}{\pi} \int_0^1
      \frac{v^2-v^{\sigma+2}}{(1-v^2)^{3/2}} dv  \right)    \right].
\label{eq:Omega_eta}
\end{equation} 
Since $\int_0^1 \frac{v^2-v^{\sigma+2}}{(1-v^2)^{3/2}} dv =
-\frac{\pi}{2}+\sqrt{\pi}\frac{\Gamma\left( (\sigma+3)/2\right)
}{\Gamma\left( (\sigma+2)/2\right)}$, the squared frequency shift is
\begin{equation}
  \delta \Omega^2=\frac{2^{\frac{\sigma+4}{2}}\beta}{\sqrt{\pi}(\sigma+2)} \frac{\Gamma
    \left(\frac {\sigma+3}{2}\right) }{\Gamma \left(\frac {\sigma+2}{2}\right)}
  \left( \frac{E}{\omega^2} \right)^{\frac{\sigma}{2}}. 
\label{eq:Omega_shift}
\end{equation} 
Eq.~(\ref{dl}) is derived from Eq.~(\ref{eq:Omega_shift}) for
$\beta=1$ and $\omega^2=\tilde{\epsilon}_l$.



\begin{thebibliography}{99}
\bibitem{particular} For $\sigma>0$ the first derivative $ | u_{l}
  |^{\sigma} u_l$, and the second derivative $ (\sigma +1) | u_{l}
  |^{\sigma}$, of the nonlinear term $| u_{l} |^{\sigma+2}/(\sigma
  +2)$, needed for the implementation of any SABA integrator with the
  addition of the corrector term C, are continuous functions. In
  particular, for the integration of the $\sigma =2$ and $\sigma=4$
  cases where we can substitute $| u_{l} |^{\sigma+2}$ by $
  u_{l}^{\sigma+2}$, we apply the SABA$_2$C integration scheme. For
  all other values of $\sigma$, which yield noninteger powers of $|
  u_{l} |$, the corrector term becomes cumbersome. For these cases the
  use of SABA$_3$ integrator was preferred over SABA$_2$C.
\end{thebibliography}
\end{document}